\documentclass{elsart}
\usepackage{graphics,amssymb}
\newcommand{\spg}{S{\mbox{p}}}
\newcommand{\bmu}{{\mbox{\boldmath$\mu$}}}
\newcommand{\bnu}{{\mbox{\boldmath$\nu$}}}
\journal{Phys. Lett. A}
\begin{document}
\begin{frontmatter}

\title{Partial positive scaling transform: a separability criterion}

\author[leb]{Olga V. Man'ko}
\ead{omanko@sci.lebedev.ru}
\author[leb]{V. I. Man'ko \corauthref{cor}}
\corauth[cor]{Corresponding author.}
\ead{manko@sci.lebedev.ru}
\author[nap]{G. Marmo}
\ead{marmo@na.infn.it}
\author[tex]{Anil Shaji}
\ead{shaji@physics.utexas.edu}
\author[tex]{E. C. G. Sudarshan}
\ead{sudarshan@physics.utexas.edu}
\author[nap]{F. Zaccaria}
\ead{zaccaria@na.infn.it}

\address[leb]{P. N. Lebedev Physical Institute, Leninskii Prospect, 53, Moscow 119991}
\address[nap]{Instituto Nazionale di Fisica Nucleare, Sezione di Napoli, Complesso Universitario di Monte Sant Angelo, Via Cintia, I-80126 Napoli, Italy}
\address[tex]{The University of Texas, Center for Statistical Mechanics, 1 University Station C1609, Austin, Texas, USA}

\begin{abstract}
The problem of constructing a necessary and sufficient condition for establishing the separability of continuous variable systems is revisited. Simon [R. Simon, Phys. Rev. Lett. {\bf 84}, 2726 (2000)] pointed out that such a criterion may be constructed by drawing a parallel between the Peres' partial transpose criterion for finite dimensional systems and partial time reversal transformation for continuous variable systems. We generalize the partial time reversal transformation to a partial scaling transformation and re-examine the problem using a tomographic description of the continuous variable quantum system. The limits of applicability of the entanglement criteria obtained from partial scaling and partial time reversal are explored.
\end{abstract}


\end{frontmatter}

\section{Introduction}\label{sec1}
The qubit - an idealized two state quantum system - is the fundamental building block out of which a universal quantum information processor may be envisioned. Any representation of a qubit on a physical system is almost always an abstraction to the extent that we ignore all except two discrete states of the quantum entity. A complete description of the physical system representing a qubit usually involves continuous degrees of freedom like positions, momenta, relative phases and so on. It is fortunate that in most cases; especially in non-relativistic settings; we can ignore the continuous degrees of freedom of the qubit and focus on a subspace of its Hilbert space spanned by the eigenstates of an operator with a finite number of discrete eigenvalues like angular momentum, spin or polarization.

Detecting, characterizing and understanding entanglement between finite dimensional quantum systems like qubits and qutrits is an important piece in the developing paradigm of quantum information processing. Entanglement \cite{schrodinger35} is considered a physical resource that is used to construct quantum algorithms that can solve computational problems that classical algorithms and classical computers cannot realistically handle. 

Given a multipartite quantum system the question is whether it is possible to find a test that will detect entanglement that may be shared between the subsystems. Since separable states form a convex subset of the convex set of all possible states of the system, the Hahn-Banach theorem assures us that such a test must exist. Identifying the test is a non-trivial problem though. For two qubit systems Peres \cite{peres96a} suggested the partial transpose criterion which states that if a two qubit density matrix goes over to another density matrix (positive, trace one matrix) under partial transposition then the original state is separable. This was subsequently shown to be a necessary and sufficient condition for detecting entanglement in qubit-qutrit systems as well as two qubit systems by the Horodeckis \cite{horodecki96a}. A necessary and sufficient entanglement criterion for arbitrary multipartite quantum systems made up of finite dimensional subsystems is still not forthcoming in spite of all the interest that the problem attracts (see, for instance \cite{special_issue02}). 

The Peres-Horodecki criterion is based on the fact that the transpose operation is a positive but not completely positive map \cite{sudarshan61a,woronowicz76a,jamiolkowski72a} on the state of a system. Choi \cite{choi72,choi74,choi75a,choi75b} has shown that a not completely positive map applied to a part of an extended system will not be positivity preserving on all states of the whole system. For the combined state of two qubits, the transpose map applied to one of the qubits (partial transpose of the two qubit state) is not positive preserving on {\em all} entangled states. The same is true for the qubit-qutrit system. For two qutrit systems and higher, there exists certain states called bound entangled states that remain positive under partial transpose even if its an entangled state and the Peres-Horodecki criterion ceases to be a sufficient condition for detecting entangled states. It is worth noting here that the transpose map applied to one of the subsystems is not a physically implementable operation. It is an algebraic manipulation that can be done on any given density matrix of the system to test for entanglement. So an alternate way of thinking about tests for entanglement is that we conceive of an operation (which may be unphysical) that makes sense only on separable states of the system and not on entangled ones. If an algebraic relation between measurable quantities of the system that encapsulates the result of this operation can be found then that can be used to test entanglement. Several partial results pertaining to higher dimensional, multipartite state developed along similar lines can be found in the literature \cite{dodunov02a,decastro03a}    

The realization that the physical representation of an elementary qubit, even as something very simple like the spin of an isolated electron, must include at least one pair of canonical variables for a complete description is motivation to find entanglement criteria for continuous variable systems. Given that two qubits represented by, say, two electrons are in a ``separable'' state as far as the spin part of their combined wave function is concerned, one would like to ask the question whether they are really in a separable state given a complete description of the system in terms of spin, position and momentum coordinates. Of course, knowing the difficulties in finding such separability criteria even for generic finite dimensional, bipartite states, one might legitimately be suspicious of the chances of finding such a criterion for bipartite systems made of infinite dimensional subsystems. 

In \cite{simon00a}, Simon pointed out that for bipartite continuous variable systems it is possible to find an analogue of the Peres-Horodecki criterion which is a necessary and sufficient test for entanglement in {\em certain restricted cases}. The partial transpose is shown to be equivalent, in the continuous variable case, to the time reversal of one of the subsystems. The partial time reversal is not only meaningful on separable states but in fact, in some cases, it is even implementable using phase conjugation. On entangled states partial time reversal ceases to be a meaningful operation though. Simon finds an algebraic relation between the second moments of canonical position and momentum operators of the bipartite system that can be used to test for entanglement based on the partial time reversal transformation. 
 
 In this paper we examine Simon's seminal result in greater detail and suggest a generalization based on partial scaling of the canonical variables. The partial time reversal is shown to be a special case of the partial scaling. We also phrase the discussion in terms of tomographic representation \cite{manko97a,manko00a,manko04a} of the state of the quantum system rather than using the Wigner function representation \cite{wigner32a,moyal49a}, the tomogram being the more readily measurable one in an experimental setting. The scaling transformation can be applied to both the Wigner function and the symplectic tomogram \cite{mancini95a,mancini96a,mancini97a} of the quantum state. The partial scaling of the tomogram has the advantage that it can be easily extended to states of systems with more than two pairs of continuous canonical variables. 

The test for entanglement devised using the partial time reversal and partial scaling transformations are based on the Heisenberg \cite{heisenberg27a} and Robertson-Schr\"{o}dinger \cite{schrodinger30a,robertson30a} uncertainty relations. The connection of the Robertson-Schr\"{o}dinger uncertainty relation to linear canonical transformations of the quadratures of quantum states with several canonical degrees of freedom is discussed in detail in \cite{dodunov80a,dodunov89a,sudarshan95a}. Here we extend the discussion to the partial scaling transformations on arbitrary bipartite states and obtain a simpler derivation of Simon's algebraic relation that can be used as a test for entanglement. We also discuss the limits of applicability of the entanglement criteria based on both the partial time reversal and partial scaling. 

The organization of this paper is as follows: In section \ref{sec2} we discuss the Robertson-Schr\"{o}dinger uncertainty relations for multi-mode states. We introduce the partial scaling transformation on tomographic representation of quantum states in section \ref{sec3}. In section \ref{sec4} we derive the criterion for entanglement based on the partial scaling transformation. The discussion in section \ref{sec5} is on the limits of applicability of the criterion and examples of systems on which the criterion will always work. Our conclusions are in section \ref{sec6}. 

\section{Uncertainty relations for composite systems} \label{sec2}

The quantum mechanical uncertainty relations expressed in terms of measurable expectation values of combinations of the canonical observables associated with a multi-mode system is the starting point for constructing a test for entanglement in such systems. We are interested in uncertainty relations that are invariant under canonical transformations of the system variables. The strategy is to then device a non-canonical transformation which, when applied to the relation, will indicate the presence of entanglement.   

Consider a system with $N$ canonical degrees of freedom with $N$ pairs of canonical variables denoted by $\xi_{\alpha}$, $\alpha = 1,2, \ldots 2N$. In our notation all $\xi_{\alpha = 2j}$ (with even indices) are variables conjugate to $\xi_{\alpha = 2j-1}$. Canonical transformations are inhomogeneous, symplectic transformations on the variables of the form;
\begin{equation}
  \label{eq:canonical1}
  \xi_{\alpha} \longrightarrow S_{\alpha \beta} \xi_{\beta} + c_{\alpha}.
\end{equation}
These transformations are elements of the group $\spg (2N, {\bf R}) \odot T(2N)$ where $\spg (2N, {\bf R})$ is the real symplectic group in $2N$ dimensions, $T(2N)$ is the group of translations and $\odot$ represents the semi-direct product between the two. Canonical transformations leave the commutation relations between $\xi_{\alpha}$ invariant . 

For one degree of freedom, identifying the bilinear invariant under canonical transformation leads to the Robertson-Schr\"{o}dinger uncertainty relation (units chosen so that $\hbar=1$): 
 \[ \xi_1 = q \quad, \quad \xi_2 = p \quad ; \quad   \langle \xi_1 \rangle  = c_1 \quad , \quad  \langle \xi_2 \rangle  = c_2\]
\begin{equation}
  \label{eq:sri}
\langle (\xi_1 - c_1)^2 \rangle \langle (\xi_2 - c_2)^2 \rangle - \left\langle \frac{\xi_1 \xi_2 + \xi_2 \xi_1}{2} - c_1 c_2 \right\rangle^2  \geq  \frac{1}{4}.
\end{equation}
Note that the usual Heisenberg uncertainty relation
\begin{equation}
\label{eq:hur}
\langle (\xi_1 - c_1)^2 \rangle \langle (\xi_2 - c_2)^2 \rangle    \geq   \frac{1}{4}
\end{equation}
is not invariant under canonical transformations \cite{sudarshan95a}. 

For simplicity in the discussion that follows we replace $\xi_{\alpha} - \langle \xi_{\alpha} \rangle$ with just $\xi_{\alpha}$ and define $Q \equiv q - \langle q \rangle$, $P \equiv p - \langle p \rangle$, $\sigma_{QQ} \equiv \langle Q^2 \rangle$, $\sigma_{PQ} = \sigma_{QP} = \frac{1}{2} \langle QP + PQ \rangle = \frac{1}{2} \langle \{ Q \, , \, P \} \rangle$ and $\sigma_{PP} \equiv \langle P^2 \rangle$. We can now write Eq. (\ref{eq:sri}) in the form 
\begin{equation}
  \label{eq:sri2}
\det C = \det \left[ V + \frac{i}{2} \Omega \right] \geq 0  
\end{equation}
where
\begin{equation}
  \label{eq:sri3}
  V = \left( \begin{array}{cc} \langle \xi_1^2 \rangle & \frac{1}{2} \langle \xi_1 \xi_2 + \xi_2 \xi_1 \rangle \\ \frac{1}{2} \langle \xi_1 \xi_2 + \xi_2 \xi_1 \rangle & \langle \xi_2^2 \rangle \end{array} \right) = \left( \begin{array}{cc} \sigma_{QQ} & \sigma_{QP} \\ \sigma_{PQ} & \sigma_{PP} \end{array} \right)
\end{equation}
and
\begin{equation}
  \label{eq:sri4}
  \Omega = \left( \begin{array}{cc} 0 & 1 \\ -1 & 0 \end{array} \right).
\end{equation}
Rewriting the Robertson-Schr\"{o}dinger uncertainty relation in terms of the dispersion matrix $V$ allows a direct generalization to systems with several canonical degrees of freedom.  

For $N$ degrees of freedom,
\begin{equation}
  \label{eq:ndeg1}
  V_{\alpha \beta}= \frac{1}{2} \langle \{ \xi_{\alpha} \, , \, \xi_{\beta} \} \rangle \quad ; \quad \alpha, \beta = 1, 2, \ldots 2N. 
\end{equation}
form a $2N \times 2N$ matrix that transforms as an irreducible second rank tensor under the linear canonical (symplectic) transformations and have $N$ invariants. The generalized uncertainty relations require that the matrix is non-negative and that each of its $2 \times 2$ minors be greater than or equal to 1/4. This condition can be written down as
\begin{equation}
  \label{eq:ndeg2}
  C_{\alpha \beta} = V_{\alpha \beta} + \frac{i}{2} \Sigma_{\alpha \beta} \geq 0
\end{equation}
where $\Sigma$ is the canonically invariant block diagonal matrix, diag$(\Omega, \Omega \ldots \Omega)$. It is easy to see that 
\begin{equation}
  \label{eq:ndeg3}
  \det V \geq \frac{1}{4^N}.
\end{equation}

While $C_{\alpha \beta}$ and $V_{\alpha \beta}$ are invariant under linear canonical transformations they are {\em not} invariant under scale changes on the $\xi_{\alpha}$ that are not contained in Sp($2N$, {\bf{R}}). In particular under scaling $C_{\alpha \beta}$ is not necessarily positive definite, much less satisfy the generalized Robertson-Schr\"{o}dinger uncertainty bounds. This property can be used to construct a test for entanglement, starting from Eq. (\ref{eq:ndeg2}). But before looking at the test for entanglement we first look at how $C_{\alpha \beta}$ can be computed for a given quantum state with continuous variables and how the scaling transformation may be implemented on it. Note that for one degree of freedom the scaled matrix $V$ is always non-negative. The determinant of $V$ being just a multiple of the determinant of the matrix before scaling. Therefore scaling will leave $C$ positive as long as the overall multiplicative factor appearing before $\det V$ is greater than unity. 

\section{Tomograms of quantum states and scaling transforms}\label{sec3}

Given a density matrix $\rho$ corresponding to a continuous variable system we can express it in the form of a distribution in phase space, $\rho (p , q)$ where,
\[ \rho(p, q) = \int \left\langle q + \frac{x}{2} \bigg{|} \rho \bigg{|} q - \frac{x}{2} \right\rangle e^{- i p x} dx. \] 
Even thought $\rho(p,q)$ appears to have continuous indices $q$ and $p$, it is in fact only of discrete countably infinite dimension by virtue of the fact that $\rho$ is in a Hilbert space. So given $\rho$ we can define a distribution in phase space. Such attempts like the Wigner-Moyal distribution given above and the diagonal coherent state representation (which is sometimes called the $P$-representation) have the problem that they are not guaranteed to be strictly non-negative. However there are two distribution functions that are strictly non-negative. The Husimi-Kano expectation values of $\rho$ in a complete set of coherent states and the quantum tomogram \cite{mancini95a,mancini96a}. The latter is defined by
\begin{equation}
  \label{eq:tomo1}
  \omega (X, \mu, \nu) = {\mbox{tr}}[\rho \, \delta (X-\mu q - \nu p)]
\end{equation}
where the operator delta distribution may be understood in terms of the Fourier integral
\begin{equation}
  \label{eq:tomo2}
  \delta(X-\mu q - \nu p) = \frac{1}{2 \pi} \int_{-\infty}^{\infty} ds \; e^{i s (X- \mu q - \nu p)}.
\end{equation}
For systems with $N$ degrees of freedom the definition of the tomogram may be generalized to
\begin{equation}
  \label{eq:tomo3}
  \omega ({\bf X} \,,\, \bmu \,,\, \bnu) = {\mbox{tr}}[\rho \, \prod_{s=1}^N \delta (X_s -\mu_s q_s - \nu_s p_s)]
\end{equation}
The tomograms are all strictly non-negative being the integrals of the Wigner-Moyal distribution along the line $\mu q + \nu p = X$. Note that even though $\omega$ appears to be the function of three variables, it is clear that the dependence on $X$ can be written as
\[ \omega(X, \mu, \nu) = \frac{1}{|X|} \omega ( 1, \mu/X , \nu /X). \]
In our discussion we retain all three labels for reasons that will become clear later.

The tomograms may be viewed as the line integrals of a density as in a classical tomogram. Since they are always non-negative, the tomograms furnish a complete set of probability functions. It can be shown that the processes of computing the tomogram from a given $\rho$ is invertible. If $\omega(X, \mu , \nu)$ are given for all $\mu$ and $\nu$ we can find a linear formula for deducing the density distribution and hence the density matrix. These are however rather complicated but still implementable (as in medical tomography). There is a direct inversion formula called the Bertrand-Bertrand formula \cite{bertrand87a,gelfand64a}.

For our present purposes we have to compute the second moments of canonical variables from the tomogram. By definition one has the tomographic dispersion matrix elements, 
\begin{equation}
  \label{eq:tomo4}
  \sigma_{X_iX_j}(\bmu,\bnu) = \int d {\bf X}  (X_i- \langle X_i \rangle) (X_j - \langle X_j \rangle) \omega ({\bf X},\bmu , \bnu) 
\end{equation}
where
\begin{equation}
\label{eq:tomo5}
\langle X_i \rangle   = \int d {\bf X} \, X_i \; \omega ({\bf X}\,,\,\bmu \,,\, \bnu).
\end{equation}
From the tomographic dispersion matrix elements we obtain the elements of $V_{\alpha \beta}$ using the following relations:
\begin{eqnarray}
  \label{eq:tomo6}
  \sigma_{Q_jQ_j} &=& \sigma_{X_jX_j}(\mu_j =1 \,,\, \mu_{i \neq j} = \nu_i =0 ) \nonumber \\
  \sigma_{P_jP_j} &=& \sigma_{X_jX_j}(\nu_j = 1 \,,\, \mu_i = \nu_{i \neq j} = 0) \nonumber \\
  \sigma_{Q_jP_j} &=& \frac{1}{2} \big[  \sigma_{X_jX_j} (\mu_j = \nu_j= 1 \,,\, \mu_{i \neq j} = \nu_{i \neq j} = 0 ) \nonumber \\
  && \hspace{1.2 cm}- \sigma_{X_jX_j} ( \mu_j =1 \,,\, \mu_{i \neq j} =  \nu_i =0 ) \nonumber \\ 
  && \hspace{1.2 cm} -\sigma_{X_jX_j}(\nu_j = 1 \,,\, \mu_i=\nu_{i \neq j} = 0) \big] \nonumber \\ 
\sigma_{Q_jP_k} &=& \sigma_{X_jX_k}(\mu_j = \nu_k=1 \,,\, \mu_{i \neq j} =  \nu_{i \neq k} = 0 ) 
\end{eqnarray}
\subsection{Scaling transform of the tomogram}\label{sec3a}
Now that we know how to compute the matrix elements of $V$ from the tomogram we look at the effects of arbitrary scaling of the canonical variables on the tomogram. 

The scaling transform
\begin{equation}
  \label{eq:scale1}
  f(q_i , p_i) \rightarrow f_S(q_i, p_i) = \int d {\bf q'} d {\bf p'} \; {\mathcal{K}}(q_i,p_i; q_i', p_i') f(q_i' , p_i' )
\end{equation}
with kernel
\begin{equation}
  \label{eq:scale2}
  {\mathcal{K}}(q_i,p_i; q_i', p_i')=\prod_{i=1}^N |\lambda_{q_i} \lambda_{p_i}| \;\delta (q_i' - \lambda_{q_i}  q_i) \delta(p_i' - \lambda_{p_i} p_i ) 
\end{equation}
induces the following transformation on the tomogram:
\begin{eqnarray}
  \label{eq:scale3}
  \omega ({\bf X}, \bmu , \bnu) & \rightarrow & \omega_S ({\bf X}, \bmu , \bnu) \nonumber \\
 & = & \int d {\bf X'} \, d \bmu ' \,d  \bnu ' \; \omega ({\bf X}', \bmu ' , \bnu')\prod_{i=1}^N \delta ( X_i - X_i') \nonumber \\
&&\hspace{1 cm} \times  \; \delta \left( \mu_i' - \frac{\mu_i}{\lambda_{q_i}} \right) \delta \left( \nu_i' - \frac{\nu_i}{\lambda_{p_i}}  \right).
\end{eqnarray}

The changes to the dispersion matrix elements brought about by the scaling transformation are given below:
\begin{eqnarray}
  \label{eq:scale4}
  \sigma_{Q_iP_j}^S & = & \frac{\sigma_{Q_iP_j}}{\lambda_{q_i} \lambda_{p_j}} \nonumber \\
\sigma_{Q_iQ_j}^S & = & \frac{\sigma_{Q_iQ_j}}{\lambda_{q_i} \lambda_{q_j}} \nonumber \\
 \sigma_{P_iP_j}^S & = & \frac{\sigma_{P_iP_j}}{\lambda_{p_i} \lambda_{p_j} }.
\end{eqnarray}

In equation (\ref{eq:tomo6}) we computed elements of the dispersion matrix $V$ by evaluating the elements of the tomographic dispersion matrix, $\sigma_{X_jX_j}(\bmu, \bnu)$ setting some of the $\mu_i$ and $\nu_i$ equal to one and all others equal to zero. The elements of $V$ after the scaling transformation is easily computed by setting the non-zero $\mu_i$ and $\nu_i$ equal to $\lambda_{q_i}^{-1}$ or $\lambda_{p_i}^{-1}$ (equal to the inverse scaling parameters of the corresponding canonical variables) rather than equal to unity.

Given a tomogram $\omega({\bf X}, \bmu, \bnu)$ we can compute the dispersion matrix $V$ and use the inequality in Eq. (\ref{eq:ndeg2}) to test if the state corresponding to the tomogram satisfies the canonical uncertainty relations or not. If the tomogram is of a physical state we are assured that the inequality is satisfied. The same protocol can be employed to test whether scaled versions of the tomogram $\omega^S({\bf X}, \bmu, \bnu)$ are acceptable to the extent that it produces dispersion matrices that respect the Robertson-Schr\"{o}dinger uncertainty relations. 

In the case of a single mode, we know that scaling of the canonical variables does not change the positivity of $V$ because the new matrix is just a positive multiple of the original one. In other words, $\det V \rightarrow \Lambda^2 \det V$. If $\omega (X, \mu ,\nu)$ is such that the matrix C computed from it satisfies $\det C = \det[V+ \frac{i}{2}\Omega] \geq 0$ then, $C$ computed from $\omega^S(X,\mu,\nu)$ will also be such that $\det C \geq 0$ for all $|\Lambda| \geq 1$. It follows that if we have a {\em separable} multi-mode state given by the tomogram
\begin{equation}
  \label{eq:sep1}
  \omega ({\bf X}, \bmu, \bnu) = \sum_k \mu_k \prod_{i=1}^N \omega^{(k)}(X_i, \mu_i, \nu_i) \; ; \; \sum_k \mu_k =1  
\end{equation}
then $\omega^S({\bf X}, \bmu, \bnu)$ obtained from arbitrary scalings of the canonical variables is always an acceptable tomogram provided $\omega({\bf X},\bmu,\bnu)$ leads to a dispersion matrix that satisfies (\ref{eq:ndeg2}).

Out of the $2N$ canonical variables $\{(q_{\alpha}, p_{\alpha}) \}, \; \alpha = 1, \ldots N$ associated with an $N$ degree of freedom system, if we change the scale of some or all of the momentum variables to $x_{\alpha} p_{\alpha} \; $ then that is not a canonical transformation. For separable states, the tomogram remains acceptable even under this partial scaling transform. For entangled states it turns out that this is not the case in general. In the next section we discuss how the partial scaling transform may be used to construct a necessary condition for the separability of multi-mode states. In restricted cases the condition is not only necessary but a sufficient test for entanglement as well. 

\section{The criterion for separability based on partial scaling} \label{sec4}

First let us look at the consequences of partial scaling on the dispersion matrix of a two mode quantum state. The state is described by the canonical variables\[ \xi_{\alpha} = ( q_1, \, p_1 ,\, q_2 \, ,p_2), \] 
with
\[ Q_i \equiv q_i - \langle q_i \rangle  \quad ; \quad P_i \equiv p_i - \langle p_i \rangle \quad ; \quad i = 1,2. \]
We change the scale of the momentum variable of the second sub-system by a factor of $x$, i.e.
\begin{equation}
  \label{eq:sca1}
  p_2 \longrightarrow  \frac{p_2}{x}
\end{equation}
The dispersion matrix $V_2$ of the second subsystem variables is transformed so that $\det V_2 \rightarrow x^2 \det V_2$. So as long as $|x| \geq 1$ we know that the partial scaling leaves the tomograms of separable two mode states acceptable (with positive $C_{\alpha \beta}$). For a generic state the transformation changes the two mode dispersion matrix so that $ V^S + \frac{i}{2} \Sigma$ is now
\begin{equation}
\label{eq:sca2}
   \left( \begin{array}{cccc} 
\sigma_{Q_1Q_1} & \sigma_{Q_1P_1} + \frac{i}{2} & \sigma_{Q_1Q_2} & x \sigma_{Q_1P_2} \\ 
\sigma_{P_1Q_1}- \frac{i}{2} & \sigma_{P_1P_1} & \sigma_{P_1Q_2} & x\sigma_{P_1P_2} \\
\sigma_{Q_2Q_1} & \sigma_{Q_2P_1} & \sigma_{Q_2Q_2} & x\sigma_{Q_2P_2} + \frac{i}{2} \\
x\sigma_{P_2Q_1} & x\sigma_{P_2P_1} & x\sigma_{P_2Q_2} - \frac{i}{2} &x^2 \sigma_{P_2P_2} \end{array} \right).
\end{equation}

Let us denote the $C$ and $V$ matrices obtained after the partial scaling by $C^x$ and $V^x$ respectively, where $x$ is the scaling parameter. To test for the separability of a two mode state we test whether $C^x= V^x + \frac{i}{2} \Sigma \geq 0$ for all values of $x$ with $|x| \geq 1$. A consequence of $V^x + \frac{i}{2} \Sigma \geq 0$ is 
\[ \det \left[ V^x + \frac{i}{2} \Sigma \right] \geq 0, \]
which reduces to 
\begin{equation}
  \label{eq:sca3}
  Ax^2 + 2Bx + C \geq 0
\end{equation}
where
\begin{eqnarray}
  \label{eq:sca4}
  A & = & \det V - \frac{1}{4} (\sigma_{Q_1Q_1}\sigma_{P_1P_1} - \sigma_{Q_1P_1}^2) \nonumber \\
  B &=& \frac{1}{4} ( \sigma_{Q_1P_2} \sigma_{P_1Q_2} - \sigma_{Q_1Q_2}\sigma_{P_1P_2}) \nonumber \\
  C &=& \frac{1}{16} - \frac{1}{4} (\sigma_{Q_2Q_2}\sigma_{P_2P_2} - \sigma_{Q_2P_2}^2).
\end{eqnarray}
For Eq. (\ref{eq:sca3}) to be always true, it is sufficient that the discriminant,
\begin{equation}
  \label{eq:sca5}
  B^2 - 4 AC \leq 0.
\end{equation}
If we write the matrix in Eq. (\ref{eq:sca2}) as a block matrix,
\begin{equation}
  \label{eq:sca6}
  V^S + \frac{i}{2} \Sigma = \left( \begin{array}{cc} V_1 & V_{12} \\ V_{12}^T & V_2 \end{array} \right) + \frac{i}{2}\left( \begin{array}{cc} \Omega & 0 \\ 0 & \Omega \end{array} \right) ,
\end{equation}
the condition (\ref{eq:sca5}) can be expressed as
\begin{equation}
  \label{eq:sca7}
 |\det V_{12} |^2 -  (4\det V - \det V_1) \left( 1 - 4\det V_2 \right) \leq 0.
\end{equation}
The inequality (\ref{eq:sca7}) must be satisfied by all separable two mode states and it need not be so for entangled ones. For certain classes of states (\ref{eq:sca7}) is satisfied {\em only} by separable ones and then the partial scaling transform furnishes a bona fide test for entanglement. We defer our discussion of this class of states to the next section and turn to the connection between partial scaling and Simon's \cite{simon00a} partial time reversal. 

The partial time reversal transformation for two mode states is a special case of the partial scaling with $x=-1$. Time reversal of one of the subsystems of a bipartite quantum system is a canonical transformation as long as the state of the system is separable. On entangled states the transformation ceases to be meaningfully defined. From (\ref{eq:sca3}), for separable states, if 
\begin{equation}
  \label{eq:sep1a}
  A + 2B + C \geq 0
\end{equation}
then 
\begin{equation}
  \label{eq:sep2}
  A - 2B + C \geq 0
\end{equation}
also. 
Combining (\ref{eq:sep1a}) and (\ref{eq:sep2}) we obtain
\begin{equation}
  \label{eq:sep3}
  \det V + \frac{1}{16} -\frac{1}{2} |\det V_{12}| \geq \frac{1}{4} (\det V_1 + \det V_2)
\end{equation}
Using
\begin{equation}
  \label{eq:sep4}
  \det V = \det V_1 \det V_2 + |\det V_{12}|^2 - {\mbox{tr}} [V_1 \Omega  V_{12} \Omega V_2 \Omega V_{12}^T \Omega ] 
\end{equation}
we can re-write (\ref{eq:sep3}) in the form
\begin{eqnarray}
  \label{eq:sep5}
  \det V_1 \det V_2 + \left(\frac{1}{4} - |\det V_{12}| \right)^2 - {\mbox{tr}} [V_1 \Omega  V_{12} \Omega V_2 \Omega V_{12}^T \Omega ] \hspace{2 cm}  \nonumber \\
\geq \frac{1}{4} (\det V_1 + \det V_2 ) 
\end{eqnarray}
which is identical to Simon's criterion for separability of two mode states obtained using partial time reversal. The partial scaling transformation provides an alternate approach to deriving the criterion. 
 
Generalizing the criterion for separability based on the partial scaling transform to systems with $N$ degrees of freedom is straightforward. Starting from the tomogram of a state of the system, $\omega({\bf X},\bmu,\bnu)$ we first verify that it satisfies the Robertson-Schr\"{o}dinger uncertainty relations by checking that $C_{\alpha \beta} = V_{\alpha \beta} + \frac{i}{2} \Sigma_{\alpha \beta}$ is a positive matrix. This may be done by making sure that all the principal minors of $C$ are positive definite. We can now perform an arbitrary scaling  described by the vector ${\bf x} = (x_1, \, x_2 , \, \ldots x_{2N})$ on the tomogram. From the scaled tomogram we can now compute
\begin{equation}
  \label{eq:gen1}
  C^{{\bf x}}_{\alpha \beta} = V^{{\bf x}}_{\alpha \beta} + \frac{i}{2} \Sigma_{\alpha \beta} \quad ; \quad \alpha, \beta = 1,2, \ldots 2N
\end{equation}
where
\begin{equation}
  \label{eq:gen2}
  V^{{\bf x}}_{\alpha \beta} = [D_{{\bf x}} V D_{{\bf x}}]_{\alpha\beta} 
\end{equation}
with $D_{{\bf x}} \equiv {\mbox{diag}}(x_1,\, x_2, \, \ldots x_{2N})$.
The $2N$ real parameters $\{x_{\alpha} \}$ parameterize the Abelian scaling semi-group and we require that 
\[ |x_1x_2| \geq 1, \; |x_3 x_4| \geq 1 , \ldots , |x_{2N-1}x_{2N}| \geq 1.\] 
The necessary condition for the separability of the state represented by the tomogram $\omega ({\bf X}, \bmu, \bnu)$ is that
\begin{equation}
  \label{eq:gen3}
  C^{{\bf x}} \geq 0
\end{equation}
for all allowed choices of ${\bf x}$.

Out of the $2N$ scaling parameters we can always choose one pair, $(x_{2k-1} ,x_{2k})$ such that $|x_{2k-1}x_{k}| = 1$ using the freedom to choose an overall scale factor that does not affect the positivity of $C^{{\bf x}}$. Furthermore, we are interested only in partial scalings that change the value of $|x_{2j -1} x_{2 j}|, \; j = 1, \ldots ,N$. Without loss of generality we can scale only the momentum variables and leave the position variables unchanged so that all $x_{2j-1}=1$ and all $x_{2j} \geq 1$. 

For two-mode systems, the choice $x_1=x_2=x_3=1, \; x_4 =x^{-1}$ exhausts all the possibilities. For three mode systems there are more choices. By choosing $x_1=x_3=x_4=x_5=x_6=1$ and $x_2=x^{-1}$ we can check whether the first mode is entangled to the remaining two. Each of the other two modes can similarly be tested for entanglement with the rest of the system. Tripartite entanglement can be tested for by scaling the momenta of two out of the three subsystems by choosing $x_1=x_3=x_5=x_6=1$ and $x_2=x^{-1}, \; x_4=y^{-1}$. The determinant of $C^{{\bf x}}$ is now a polynomial in both $x$ and $y$ which is positive semi-definite in some domain in the $xy$-plane with $|x|,|y| \geq 1$.  

\section{Discussion}\label{sec5}
The scaling transformation is not a canonical transformation and can be thought of as an effective scaling of the Planck's constant. A positivity of the density operator of a separable state is not sensitive to such scalings applied to individual sub-systems. The entangled states, on the other hand, are sensitive to such scalings and in many cases this shows up by making $C^{{\bf x}}$ negative for certain choices of ${\bf x}$. 

Positivity of $C^{{\bf x}}$ is a condition involving {\em only} the second moments of symmetrized combinations of the canonical operators associated with the system. For distributions of continuous variables we require {\em  all} the moments if we are to have complete information about the distribution. So positivity $C^{{\bf x}}$ does not guarantee that the distributions from which it came from corresponds to a separable state unless the second moments determine all the higher moments and the distribution itself. 

For symmetric distributions, all odd moments vanish and for a class of distributions of which the most familiar is the Gaussian distribution, the second moment determines all the higher moments. So for systems characterized by Gaussian distributions the positivity of $C^{{\bf x}}$ under partial scalings become a necessary and sufficient condition for separability. 

The construction of the matrix $C^{{\bf x}}$ is related to the $f-$oscillator construction introduced in \cite{manko97b}. For $f-$oscillators the deformed annihilation operator has the form
\begin{equation}
  \label{eq:fosc1}
  A = a f(a^{\dagger}a)
\end{equation}
where $a^{\dagger}$ and $a$ are the standard creation and annihilation operators. If the function $f(a^{\dagger}a)$ is a constant and equal to $\lambda^{-2}$ then the non-canonical transformation (\ref{eq:fosc1}) corresponds to the Planck's constant being scaled to $\lambda \hbar$. Thus one can formulate the response of the density operator to scaling of the canonical variables in terms of the $f-$deformed oscillator with the function $f(a^{\dagger}a)$ equal to a constant. 

\subsection{Limits of applicability of the criterion} \label{sec5a}

The criterion for separability can fail to detect entanglement in a mixture of two Gaussian distributions even if it is a necessary and sufficient condition on a single Gaussian. We construct an example of such a situation in this section, illustrating the limits of applicability of the partial scaling and partial time reversal transforms as the basis for constructing tests for entanglement.

Consider a simple bipartite pure Gaussian state given by the wave function
\begin{equation}
  \label{eq:gauss1}
  \Psi_1({\bf q})= \frac{1}{\sqrt{\pi \sqrt{|M|}}}\exp\left[{-\frac{1}{2} {\bf q}^T \cdot M^{-1} \cdot {\bf q}} \right] 
\end{equation}
with
\[ {\bf q} \equiv \left( \begin{array}{c} q_1 \\ q_2 \end{array} \right) \quad , \quad M = \left( \begin{array}{cc}m_{11} & m \\ m &  m_{22} \end{array} \right)
\]
and 
\[ |M| = \det M = m_{11}m_{22}-m^2 \geq 0. \]
We have assumed for simplicity that all the elements of $M$ are real. Since we assume for the purposes of the following analysis that we {\em know} the state, we may choose to describe it in terms of its tomogram, Wigner function or any other possible representation. The computation of the second moments of the distribution is transparent if we represent the state using its Wigner function, so we choose this option. The Wigner function corresponding to the Gaussian state is
\begin{eqnarray}
  \label{eq:gauss2}
  W({\bf q} \,,\, {\bf p}) &=& \frac{1}{4 \pi^2} \int d^2{\bf x} \; \Psi^*({\bf q + x/2})\Psi({\bf q - x/2}) e^{-i {\bf p \cdot x}} \nonumber \\
&=& \frac{1}{\pi^2} \exp \left[ -({\bf q}^T \cdot M^{-1} \cdot {\bf q} + {\bf p}^T \cdot M \cdot {\bf p}) \right]. 
\end{eqnarray}
Using the Wigner function we can compute the non-zero second moments of the state:
\begin{equation}
  \label{eq:gauss3}
  \langle q_1^2 \rangle = \frac{m_{11}}{2} \;,\;  \langle q_2^2 \rangle = \frac{m_{22}}{2} \;,\;  \langle q_1 q_2 \rangle = \langle q_2 q_i \rangle = \frac{m}{2} 
\end{equation}
and
\begin{equation}
  \label{eq:gauss4}
\langle p_1^2 \rangle = \frac{m_{22}}{2 |M|} \;,\;  \langle p_2^2  \rangle = \frac{m_{11}}{2 |M|} \;,\;  \langle p_1 p_2 \rangle = \langle p_2 p_1 \rangle = - \frac{m}{2 |M|}   
\end{equation}
The Robertson-Schr\"{o}dinger uncertainty relation for the state is given by 
\begin{equation}
  \label{eq:gauss5}
  C =  \frac{1}{2} \left( \begin{array}{cccc} 
m_{11} & i & m & 0 \\
-i & \frac{m_{22}}{|M|} & 0 & -\frac{m}{|M| }\\
m & 0 & m_{22} & i \\
0 & -\frac{m}{|M|} & -i & \frac{m_{11}}{|M|} \end{array} \right) \geq 0  
\end{equation}
The pure Gaussian state in (\ref{eq:gauss1}) is a minimum uncertainty state and we find that $\det C =0$ as expected. Partial scaling, $p_2 \rightarrow x^{-1}p_2$ results in 
\begin{equation}
  \label{eq:gauss6}
  \det C^x = -(x-1)^2\frac{m^2}{16 |M|} \leq 0
\end{equation}
which shows that the state is indeed entangled.

Now consider another pure Gaussian state
\begin{equation}
  \label{eq:gauss7}
  \Psi_2({\bf q}) = \frac{1}{\sqrt{\pi \sqrt{|N|}}}\exp\left[{-\frac{1}{2} {\bf q}^T \cdot N^{-1} \cdot {\bf q}} \right] 
\end{equation}
with
\[N = \left( \begin{array}{cc}n_{11} & n \\ n &  n_{22} \end{array} \right) \quad, \quad n_{11},\, n_{22},\, n \in {\bf R}.\]

A mixed state constructed out of the pure states in (\ref{eq:gauss1}) and (\ref{eq:gauss7}) of the form,
\begin{equation}
  \label{eq:gauss8}
  \rho = \alpha \rho_1 + (1-\alpha) \rho_2 \quad ; \quad \rho_i \equiv \Psi_i \Psi_i^{\dagger} \quad , \quad 0 \leq \alpha \leq 1
\end{equation}
also satisfies the Robertson-Schr\"{o}dinger inequality. The Wigner function for the mixed state reads, 
%
%
\begin{eqnarray}
  \label{eq:gauss8a}
  W({\bf q}\,,\, {\bf p})  &= & \frac{\alpha}{\pi^2}  \exp \left[ -\frac{m_{22}}{|M|} q_1^2 - \frac{m_{11}}{|M|}q_2^2 -m_{11}p_1^2 -m_{22}p_2^2 \right. \nonumber \\ 
&& \hspace{6 cm} \left. + 2\frac{m}{|M|}q_1q_2  -2m p_1p_2 \right]   \nonumber \\
& & +   \frac{1-\alpha}{\pi^2}  \exp \left[ -\frac{n_{22}}{|N|} q_1^2 -\frac{n_{11}}{|N|}q_2^2 - n_{11}p_1^2 - n_{22}p_2^2 \right. \nonumber \\
&& \hspace{6 cm} \left. + 2\frac{n}{|N|}q_1q_2  -2n p_1p_2 \right] 
\end{eqnarray}
which, by inspection, is a non-separable function of the two pairs of canonical variables. The Robertson-Schr\"{o}dinger uncertainty relations for the state is given by the inequality $C_{mix} \geq 0$ where $C_{mix}$ is the matrix:
\[
\hspace{-0.7 cm} \frac{1}{2} \left( \begin{array}{cccc} 
      \alpha m_{11}+ (1-\alpha) n_{11} & i & \alpha m + (1-\alpha) n & 0 \\
      -i & \frac{ \alpha m_{22}}{|M|} + \frac{(1-\alpha)n_{22}}{|N|} & 0 & - \frac{\alpha m}{|M|} - \frac{(1-\alpha)n}{|N|}  \\
      \alpha m + (1-\alpha) n & 0 & \alpha m_{22} + (1-\alpha) n_{22} & i \\
      0& - \frac{\alpha m}{|M|} -  \frac{(1-\alpha)n}{|N|} & -i &  \frac{\alpha m_{11}}{|M|} +  \frac{(1-\alpha)n_{11}}{|N|}
      \end{array} \right).
\]

We find that
\begin{equation}
  \label{eq:gauss10}
  \det C_{mix} = \frac{\alpha^2 (1-\alpha)^2 (\det [M-N])^2}{16 |M| |N|} 
\end{equation}
which is positive. Finding the determinant of the matrix $C^x_{mix}$ after an arbitrary scaling $p_2 \rightarrow x^{-1} p_2$ has been applied is arduous and hardly enlightening but in the special case where $x=-1$ when the partial scaling reduces to the partial time reversal we obtain
\begin{equation}
  \label{eq:gauss11}
  \det C^x_{mix} = \det C_{mix} - \frac{\alpha m + (1-\alpha)n}{4} \left[ \frac{\alpha m }{|M|} + \frac{(1-\alpha)n}{|N|} \right].
\end{equation}
We see that as long as
\[ \frac{\alpha m + (1-\alpha)n}{4} \left[ \frac{\alpha m }{|M|} + \frac{(1-\alpha)n}{|N|} \right] \leq \det C_{mix}\]
the mixture of two entangled states appear to be separable with respect to the partial time reversal criterion. To view this as a {\em failure of the test} for entanglement one needs alternate ways of showing that the mixed state for which the Robertson-Schr\"{o}dinger uncertainty relations are satisfied even after partial time reversal is in fact an entangled state. Since no such alternate method of showing that the state is entangled is forthcoming (otherwise we could just as well have used that as a test for entanglement) we look at a particular choice of $M$, $N$ and $\alpha$ for which the test seems to fail and see what we may conclude from it. 

\begin{figure}[!ht]
 \resizebox{8 cm}{5 cm}{\includegraphics{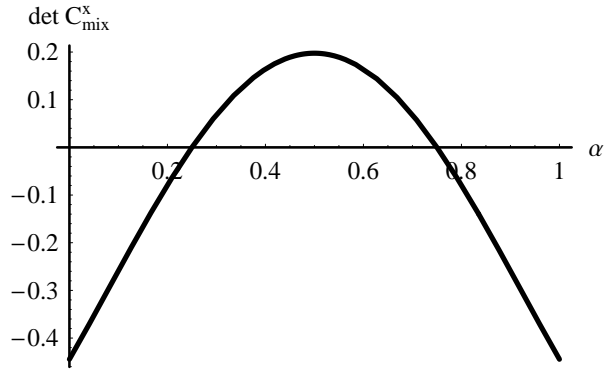}}
 \caption{$\det C^x_{mix}$ for $x=-1$ with $m_{11}=n_{11}=m_{22}=n_{22} = 0.5$ and $m=-n=0.4$.} \label{fig1}
\end{figure}

The choice $m_{11}=n_{11}$, $m_{22}=n_{22}$ and $m=-n$ (so that $|M| = |N|$ but $M \neq N$) simplifies Eq. (\ref{eq:gauss11}) to
\begin{equation}
  \label{eq:gauss12}
  \det C^x_{mix} = \frac{m^4 \alpha^2 (1-\alpha)^2}{|M|^2} - \frac{m^2 (2 \alpha -1)^2 }{4 |M|} 
\end{equation}
Setting $\det C^x_{mix} =0$ we obtain a fourth order equation in the parameter $\alpha$. If we can find at least one solution $\alpha_1$ in the open interval $(0,1)$ we know that for a finite range of values of $\alpha$ between zero and one the scaled second moment matrix is such that the Robertson-Schr\"{o}dinger uncertainty relations are satisfied. The equation $\det C^x_{mix} =0$ can be solved analytically using the change of variables $\alpha \rightarrow \frac{1}{2} - \beta$ reducing it to an equation in even powers of $\beta$ with solutions
\[ \beta^2 = \frac{1}{4} \left( 1 + \frac{2 |M| \pm \sqrt{|M|(|M|+m^2)}}{m^2} \right). \]
A plot of $\det C^x_{mix}$ as a function of $\alpha$ for a particular choice of $M$ and $N$ is given in Fig. \ref{fig1}.

From Eq. (\ref{eq:gauss8a}) we know that for generic values of $\alpha$ the state is not separable. On the other hand from Fig. \ref{fig1} we see that for a continuous range of values of $\alpha$ the partial time reversal (and partial scaling) criterion suggests that the state is separable. In the example we considered, by choosing $m=-n$, we are engineering the state so that the contribution to the cross correlations $\langle q_1 q_2 \rangle$ and $\langle p_1 p_2 \rangle$ from each of the two pure components of the state cancel each other. This is the reason why our test for entanglement based only of the second moments of the distribution fails. For generic values of $\alpha$ the mixed state we considered is entangled by construction.  Since the sum of two Gaussian states is not another Gaussian we do not expect the second moments to determine all higher moments. We therefore expect signatures of the entanglement to show up in the nature of higher moments of the distribution. 
 
\section{Conclusion} \label{sec6}

We have derived a necessary (and in some cases sufficient) condition for the separability of quantum states of a multi-mode system with continuous variables. The criterion is based on the partial scaling transform of the canonical variables of the system. We show how the scalings may be implemented on tomographic descriptions of the states. The test for separability is formulated in terms of the positivity conditions on the matrix of second moments computed from the tomogram. These conditions are based on the Robertson-Schr\"{o}dinger uncertainty relations that all physical states must satisfy. Under partial scaling the matrix of second moments behaves differently for separable and entangled states because partial scaling transforms a state to another physical state only if it is separable. The criterion for separability obtained using the partial scaling transformation reduces to the familiar result due to Simon \cite{simon00a} for separability of bipartite states when the scaling corresponds to a partial time reversal. 

For those states represented by distribution functions like the Gaussian for which the second moment determines all the features of the distribution, the criterion derived here is a necessary and sufficient test for separability. We have also shown that the test is limited in its applicability because of the fact that it is based solely on the second moments and therefore cannot depend on all the features of the overall state.  

\section*{Acknowledgments}

O. V. M is grateful to the Russian Foundation for Basic Research for partial support under project no. 03-02-16408

V. I. M. and E. C. G. S. thank Dipartimento di Scienze Fisiche, Universita ``Federicl II'' di Napoli and Istituto Nazionale do Fisica Nucleare, Sezione di Napoli for kind hospitality

A. S. acknowledges the support of US Navy - Office of Naval research through grant Nos. N00014-04-1-0336 and N00014-03-1-0639

\bibliographystyle{elsart-num}
\bibliography{criterion,ncp}

\end{document}